\pdfoutput=1

\documentclass{svproc}

\usepackage{url}
\usepackage{multirow}
\usepackage{amsmath}
\usepackage{amssymb}
\usepackage{adjustbox}

\usepackage[sort&compress,numbers]{natbib}
\usepackage{booktabs}
\usepackage{longtable}
\usepackage{array}

\usepackage{hyperref}
\hypersetup{
	colorlinks   = true, 
	urlcolor     = blue, 
	linkcolor    = blue, 
	citecolor   = blue 
}

\begin{document}
\mainmatter              

\title{A Novel Mathematical Model for Infrastructure Planning of Dynamic Wireless Power Transfer Systems for Electric Vehicles}

\titlerunning{Dynamic Wireless Power Transfer Systems}  

\author{Afshin Ghassemi\inst{1} \and Laura Soares\inst{2} \and Hao Wang\inst{3} \and Zhimin Xi\inst{4}
}

\authorrunning{A. Ghassemi, L. Soares, H. Wang, and Z. Xi} 

\tocauthor{xx and xx}
\institute{Rutgers University, Piscataway NJ 08854, USA, \email{ag1898@soe.rutgers.edu}
\and
Rutgers University, Piscataway NJ 08854, USA, \email{lm804@soe.rutgers.edu}
\and
Rutgers University, Piscataway NJ 08854, USA, \email{hw261@soe.rutgers.edu}
\and
Rutgers University, Piscataway NJ 08854, USA, \email{zx112@soe.rutgers.edu}}
\maketitle

\begin{abstract}
About 26\% of total U.S. energy consumption is used in the transportation sector. Conventional vehicles use fuels such as gasoline, emit harmful gases, and have adverse effects on the environment. Electric vehicles (EVs) provide an alternative solution that decreases the dependency on traditional fuels such as gasoline and reduces hazardous gas emissions. EVs can drive longer distances by employing dynamic wireless power transfer systems (DWPT) without increasing their battery size or having stopovers. Additionally, developing a decision system that avoids an excessive load on the power grid is essential. These decision systems are particularly beneficial for autonomous driving for personal and public transportation. This study briefly reviews the available literature in dynamic wireless power transfer systems and proposes a novel system-level mathematical decision model to find the optimal profile for wireless charging infrastructures. We analyze the role of renewable energy integration on DWPT systems and identify the framework, benefits, and challenges in implementing DWPT for EVs. The mathematical model is mixed-integer, multi-period, and linear, minimizing the total system cost while satisfying the systems requirements. The proposed model and the case study analysis in this research determine the near-optimal plan for DWPT infrastructure allocations and pave the road toward a more detailed power grid and renewable energy integration. Our result indicates that renewable energies can significantly decrease the DWPT total system cost, infrastructure requirements and increase the EVs’ reliability. 
\end{abstract}

\section{Introduction}

The traditional transportation sector uses more than 26\% of all the generated energy in the USA. It is highly dependent on conventional energy sources such as gasoline, accounting for 56\% of total transportation energy consumption in the USA. Gasoline usage can cause adverse environmental effects and increase greenhouse gases \cite{EIA2021}. The tailpipe emission also causes harmful effects on human health \cite{article2}. Electric vehicles (EVs) and Plug-in Hybrid Electric Vehicles (PHEV) are considered alternative solutions to traditional gasoline-based vehicles. All the EVs include a battery that can be charged in one of the two main methods: 
\begin{enumerate}

    \item Conductive power transfer (connected method), which requires physical contact between the vehicle and the charger. This technique can be a plug-in which can generally take between 6-8 hrs \cite{8353730}, commonly used in EV charging stations, and can charge a vehicle from 20 hours (Level 1) to 20 minutes when using a DC fast charging station \cite{HE2019452,bansal_2015}. Conductive charging also includes overhead, road-bound, or roadside charging. The vehicle is equipped with a connector device linked with supply lines charging the battery in stationary mode or dynamically \cite{wevj9010009}. In this method, vehicles cannot move during the charging process, but it is generally more efficient than Dynamic Wireless Charging (DWC).
    \item Wireless Power Transfer (WPT) charges the battery using inductive technology by embedding a charging unit with a resonant coil system on the pavement that transfers power to a secondary coil installed in the vehicle. The main advantage of WPT is that it can be implemented in both stationary and dynamic settings. The convenience of this method is that there is no need to stop the vehicle while charging. This feature opens the door for fully autonomous EVs, extends the driving range, saves the energy network more space, which is valuable in dense urban areas \cite{Song16}, helps to reduce the size of the EV's battery, is visually more aesthetic \cite{su13031270}, decreases driver's range anxiety, has minimal impact on the traffic, and is free from damage and vandalism when embedded on the pavement. There is also another similar wireless charging method when the EV decelerates to or accelerates from a stopping point, and it is called Quasi Dynamic Wireless Charging (QDWC). More details about this charging method can be found in \cite{7878666}.
\end{enumerate}

Stationary wireless charging is safer and more reliable in comparison with DWC, when the EVs are moving \cite{7731966}. Both static and dynamic wireless charging methods need almost the same charging time and logistic requirements. However, dynamic wireless charging has caused new infrastructural, design, and operational challenges that scientists in this area are trying to address.

The first DWC vehicle was made in Korea Advanced Institute of Science and Technology in 2009 \cite{JANG2018844}. Subsequently, significant portions of the research have been focused on the technical aspects of the wireless charging technology for electric vehicles \cite{9681af48fb114fe3b70035c6940db613}. Those include areas such as power flow control \cite{7110445}, and improvements in the efficiency and reliability for wireless charging pads \cite{5709980}. For example, \cite{8063878} introduces a system design that increases the expected value of transferred energy by more than 30\% in comparison to the traditional wireless charging methods for EVs.

One of the gaps in the DWPT literature is the lack of systems-level study to integrate all the agents involved in this interdisciplinary, complex system \cite{mohamadi2020nash}. The dynamic wireless power transfer system includes decisions from sub-systems such as: 

\begin{itemize}
\item Economical and environmental cost-benefit analysis. 
    \item Technological standards, safety, and regulations.
    \item Infrastructure planning.
    \item Traffic and electrified road design.
    \item City and infrastructure planning.
    \item Battery design.
    \item Wireless technology.
    \item Smart grid planning.
    \item Renewable energy integration planning.
    \item Billing and pricing optimization.
\end{itemize}

Each of these sub-systems also has operational, tactical, and strategical decision layers embedded. As a result, it is critical to perform more in-depth systems-level studies to integrate and analyze all these sub-systems. 
If we generate decisions for each sub-system in an isolated environment and neglect its relation to other sub-systems, the model decisions would be non-optimal, and the model could even become infeasible. One crucial factor to consider when we develop a systems-level study is to have a clear definition for the scale and the scope of the system; otherwise, we will face an over-complicated model. This case is especially true about DWT systems. Explicit assumptions and precise determinations of the system's boundaries can provide us with accurate decisions. 

In this study, after a brief review of different sub-categories in DWC, we propose a novel, systems-level, mathematical model that includes various sub-systems. We apply a numerical case to illustrate how a systems-level study is critical for future work on DWC planning. One of the first steps to develop a WPT for EVs is to evaluate if such a system has a significant economic and environmental value. For example, feasibility studies with regard that have been done in Europe \cite{7110445} and the US \cite{8930076} for major roads that result in reliable operation, unlimited driving range, and no recharge downtime, with a reduction in the on-board battery (36\%) and more road coverage (69\%), at the minimized total cost. Another area where WPT systems are lacking is regulation, standards, and policy. Some state-of-the-art research in this area (see for example \cite{Brecher2014ReviewAE}) aims to prepare the early standards and regulations for WPT systems.

There are already extensive literature reviews to study wireless charging for EVs \cite{osti_1399989,Gil2014ALR}. Due to the interdisciplinary nature of WPT systems, each of these studies overviews different layers of WPT literature. They cover technological and infrastructural challenges to develop a DWPT system, focusing on high-power WPT, power transmission systems, and finally, extending battery life \cite{doi:10.1177/0020294019885158,9177076,9219241}.

The elements of wireless power transfer systems for EVs are as following: \begin{enumerate}
    \item \textbf{Power source}: A power source can be connected to the grid or/and to an integrated renewable power source, such as a photovoltaic or wind system.
    \item \textbf{Power transmission system}: The power transmission system transfers the power via the electromagnetic field. The power transfer is based on Faraday’s and Lenz’s law, in which when a time-variant current is applied in a conductor, a magnetic field is created around it \cite{Ahmad2018ACR}.
    \item \textbf{Pickup system}: The pickup system receives the power by the induced current on the secondary conductor due to the time-variant magnetic flux \cite{Ahmad2018ACR}, then transmits it to either the battery or motor of the vehicle. 
\end{enumerate}

Several studies specifically aim to review the literature of DWPT systems \cite{Majhi2020ASR,Ahmad2018ACR,Machura2019ACR}. Each of these studies covers part of this interdisciplinary problem. Since we propose a location planning optimization model for wireless charging infrastructures, we provide a more detailed review of the infrastructure allocation studies in the next section.  
\section{Infrastructure Allocation}
There are many benefits associated with WPT, including impact on the battery size, driving range, and environmental effects; however, the construction is costly. Therefore, a WPT project needs to have the chargers' locations and size optimized. The studies in this area can be divided into two main categories:
\begin{enumerate}
    \item The first category is aiming at an operational network-level analysis of WPT systems. In this category, studies primarily focus on a limited number of EVs and concentrate on finding the best settings for Ev's battery, the best locations for wireless charging pads, and the best routes for the EVs. An example of the first category is planning for a group of public transit buses. Since the scope of the problem is very limited, all the needed data, such as detailed traffic conditions or EV's energy consumption, are accessible (see: \cite{JANG2015222,KO201567}. 
These types of study for DWPT are integrated into static plug-in models \cite{SUN2020102331}. 
\item The second category has a broader vision that tries to focus on the systems level. As a result, when researchers design a decision model for this category, it is infeasible usually, and they have no other choice than to use approximate approaches \cite{JANG2018844}. 
\end{enumerate}

Different mathematical techniques, such as optimization models, are utilized to model the problem in both categories \cite{CHEN2016344,CHEN2017185,7837718}.

There is a range of objectives for DWPT infrastructure allocations, such as minimizing total travel time, minimizing total system energy consumption, minimizing air pollutants and the Greenhouse gases footprint, or minimizing total system cost \cite{NGO2020102385}. These studies are implemented on a small scale for a real-world case, such as locating DWC in a network consist of selected highways in California  \cite{FULLER2016343}.
The authors developed an optimization model aiming to minimize the capital cost of the dynamic DWPT infrastructure while providing enough energy for an EV to travel between key cities within the state. Fuller \cite{FULLER2016343} overcame limitations of previous studies using a link-based formulation, different from the node-based that is an invalid approach for dynamic charging. Liu and Wang \cite{article1} developed a model to define the location of different types of charging facilities such as dynamic, stationary wireless, and plug-in, to minimize the public social cost while not exceeding a pre-determined budget \cite{RePEc:eee:transb:v:103:y:2017:i:c:p:30-55}. This study’s novelty approach is to include the driver’s route choice behavior. A multi-class user equilibrium is developed to describe the routing choice, assuming that the drivers would select the path with minimum travel cost. 

Chen et al. \cite{su12218971} used a meta-heuristic  approach of Genetic Algorithm (GA) to optimize wireless charging locations for electric bus transit systems. The study aimed to optimize costs, greenhouse gas emissions, battery service life and capacity size, and the number of charging infrastructure. 
The developed model simultaneously optimized DWPT locations and battery capacity. The results were compared with the traditional stationary charging at the bus terminal system, and they found that the DWPT is a better alternative for both cost and emission reduction.
The study tried to optimize costs, greenhouse gas emissions, battery service life, battery capacity size, and the number of charging infrastructure. 

A WPT project encompasses multiple variables, including battery specifications, charger specifications, road network, and traffic characteristics, power grid limitations, and budgetary constraints. The EV’s battery capacity and the initial state of charge (SOC) are essential factors in designing the system. The battery capacity accounts for the maximum amount of energy stored, while the SOC is the level of charge relative to the capacity. Limiting the charging power to the battery capacity is not an accurate assumption since the battery’s SOC range decreases with time and usage. The charger’s specifications, such as the number of WPT facilities, charger power, size, and efficiency with vehicle speed data, provide the information to calculate the energy transferred. The required energy will be defined by the project constraints, which could be budgetary cost \cite{RePEc:eee:transb:v:103:y:2017:i:c:p:30-55,article5,article}, trip completion assurance \cite{CHEN2017185}, or range assurance \cite{article1}. Previous models were developed to optimize capital cost \cite{FULLER2016343}, travel costs\cite{RePEc:eee:transb:v:103:y:2017:i:c:p:30-55}, range anxiety \cite{article5}, mobility or traffic flow \cite{article,Riemann2015OptimalLO}, and social cost \cite{CHEN2016344}. Although a portion of the model inputted data will be obtained by battery specifications, charger specifications, traffic data, and road network design, researchers face uncertainties that need to be addressed to build the model.

Ignoring the uncertainty in both categories leads to disastrous consequences. Thus, researchers aim to address the uncertainty for the key parameters such as travel time, energy consumption, and power grid energy availability. The uncertain approaches have been widely used for static power transfer systems \cite{SHI20201067,7815433,CAO2020105628,WANG2020119886}. More recently, probabilistic approaches are utilized for small networks for DWC systems \cite{LIU201777}. 
Liu et al offered a robust optimization approach of location wireless charging facilities for electric buses. The probabilistic approach not only can be used in location planning but on other areas such as route optimization for EVs based on DWC \cite {8402042}.

The model proposed in this study is developed in a flexible way to be conveniently extended to an uncertain model. It is also ready to be integrated for renewable energy and smart energy integration.

The following section proposes a novel, data-driven systems-level decision model to find optimal locations and settings for wireless charging pads. To our knowledge, none of the studies in this area have proposed a decision model on this scale.  

\section{System Architecture, Assumptions, and Limitations}
A model based on the network theory is proposed to find the optimal locations and lengths of the wireless charging infrastructure. 
Hourly traffics, wireless power transfer systems, average charging flows are modeled. The model allows for the flow of EV's and traditional Vehicles to move between different routes. The average charging flow of EVs is affected by: 
\begin{itemize}
\item Their initial charging level.
\item Electricity consumption by the EVs.
\item Increase in charging level due to the potential DWP systems.
\end{itemize}

The decision model aims to minimize the total system cost while not violating the minimum desired average charging flow. 
The model has the following three assumptions: 
\begin{itemize}
\item The wireless charging pads will not change the traffic behaviors of EV drivers.
\item The average flow in routes at a time period (e.g., hourly periods) of planning affects the average flow of all other connected routes at that planning time period. 
\item The charging levels for the average flow only change at intersections in each time period.
\end{itemize}
The model developed is a mixed-integer, multi-period, linear model developed to generate tactical and strategical decisions. In the next section, the deterministic optimization model is described in detail.
\section{Optimization Model}
The goal of the optimization model is to minimize the total system costs that include the construction cost of wireless charging infrastructures and the fixed and variable cost of wireless charging for all the routes that are studied.
The decision considered in the model include:
\begin{enumerate}
    \item Which locations out of a set of potential locations for wireless charging pads lead to a more reliable system and lower total system cost?
    \item What would be the best lengths for wireless charging pad lanes from a set of potential lengths?
\end{enumerate}
Continuous variables are used to represent the average charging flows, and binary variables are used to describe the control variables.

Index sets, parameters, and decision variables are described in Table~\ref{tab:parameters}.
\begin{center}
\begin{longtable}[c]{p{0.9in}|p{3.7in}} 

\caption{Sets, parameters, and variables in the proposed model}
\label{tab:parameters}\\ 
 \hline\hline 
 \endfirsthead
  \hline
  \multicolumn{2}{c}{Continuation of Table \ref{tab:parameters}}\\
 \endhead
  \multicolumn{2}{c}{Continued on next page}\\
  \hline
 \endfoot
  \hline
 \multicolumn{2}{c}{End of Table}\\
 \hline\hline
 \endlastfoot

\textbf{Sets} &  \\ \hline 
\textit{I, (J alias)} & Set of nodes/intersections that \textit{i}= $\mathrm{\{}$1, 2,{\dots}, \textit{I}$\mathrm{\}}$ \\ \hline 
\textit{T} & Hourly planning horizon that \textit{t}= $\mathrm{\{}$1, 2,{\dots}, \textit{T}$\mathrm{\}}$ \\ \hline 
\textit{M} & Set of potential locations for charging pads \textit{m}= $\mathrm{\{}$1, 2,{\dots}, \textit{M}$\mathrm{\}}$ \\ \hline 
\textit{L} & Set of potential lengths for charging pads \textit{l}= $\mathrm{\{}$1, 2,{\dots}, \textit{L}$\mathrm{\}}$ \\ \hline 
\textit{N} & Set of \textit{i} and \textit{j} where there is a route form \textit{i} to \textit{j} \textit{l}= $\mathrm{\{}$1, 2,{\dots}, \textit{N}$\mathrm{\}}$ \\ \hline 
\textbf{Parameters} &  \\ \hline 
${CC}_{ijt}$ & Average charging consumption for route \textit{i} to j at time \textit{t} \\ \hline 
${AHD}_{ijt}$ & Hourly traffic flow for route \textit{i} to \textit{j} at time \textit{t} \\ \hline 
${CCV}_{ijmt}$ & Variable cost of wireless charging for route \textit{i} to \textit{j} at location \textit{m,} time \textit{t}  \\ \hline 
${CCF}_{ijmt}$ & Fixed cost of wireless charging for route \textit{i} to \textit{j} location \textit{m} time \textit{t} \\ \hline 
${WE}_{ijt}$ & Wireless charging efficiency for route \textit{i} to \textit{j} at time \textit{t (is a function of }${AHD}_{ijt}$\textit{)} \\ \hline 
${CCC}_{ijml}$ & Construction cost of charging pads for route \textit{i} to \textit{j} at location \textit{m} with the length of \textit{L} \\ \hline 
$BC$ & Budget constraint \\ \hline 
$W_{ijt}$ & Weights of the flow for route \textit{i} to \textit{j} \textit{(calculated by }${AHD}_{ijt}$\textit{)\newline Note=0 if there is no route from I to j} \\ \hline 
$WC_{ijmlt}$ & Average charging flow to wireless charging at route \textit{i} to \textit{j} from charging pad \textit{m} with the length of \textit{l} at the time \textit{t} \\ \hline 
${UO}_{jt}$ & Average out of the system flow at node \textit{j} at time \textit{t} \\ \hline 
${WO}_i$ & Weights of the flow for outside the system route to i based on AADT \\ \hline 
$MA$ & Minimum acceptable average charging level for all routes \\ \hline 
\textit{M} & A sufficiently large number \\ \hline 
\textbf{Variables} &  \\ \hline 
$x_{ijml}$ & Binary variable to indicate whether the potential location \textit{m} for route \textit{i to j} is selected for adding charging pads with the length of \textit{l}. \\ \hline 
$u_{ijt}$ & Average charging level at route \textit{i} to \textit{j }at time \textit{t} \\ \hline 
$p_{ijt}$ & Binary variable to show the on and off state of the charging pads \\ \hline 
$z_{ijt}$ & Supplementary variable to be used in the linearization process \\ \hline 
\end{longtable}
\end{center}
\begin{equation} \label{eq:1}
\begin{array}{c}
\/Min Z:\ \sum^I_i{\sum^J_j{\sum^M_m{\sum^L_l{{CCC}_{ijml}}\times x_{ijml}}}}\\ +
\sum^I_i{\sum^J_j{\sum^M_m{\sum^L_l{\sum^T_t({{CCF}_{ijmt}\times x_{ijml}+ {CCV}_{ijmt}\times }}}}WC_{ijmlt}\times x_{ijml}}\times p_{ijt})\
\end{array}
\end{equation}
\textbf{}

\textbf{Subject to:}

\begin{equation} \label{eq:2}
\begin{array}{c}
\/u_{ijt}=(\sum^K_k{W_{kit}}\times u_{kit}+ \ {WO}_i\times {UO}_{it}) /\ (\sum^K_k{W_{kit}}+{WO}_i)\ \\-{CC}_{ijt}+\left(\sum^M_m{\sum^L_l{WC_{ijmlt}\times \ x_{ijmlt}}}\right)\ \ \ \ \ \forall \ t, (i,j)\in{N}\ 
\end{array}
\end{equation}

\begin{equation} \label{eq:3}
\begin{array}{c}
u_{ijt}\ge MA\ \ \ \ \ \ \ \forall \ t, (i,j)\in{N}\
\end{array}
\end{equation}

\begin{equation} \label{eq:4}
\begin{array}{c}
\sum^L_l{x_{ijml}\le 1\ \ \ \ \ \forall \ i,j},m
\end{array}
\end{equation}
\begin{equation} \label{eq:4.5}
\begin{array}{c}
\sum^L_l{x_{ijml}\le 0\ \ \ \ \ \forall \ m, (i,j)\notin{N}}
\end{array}
\end{equation}
\begin{equation} \label{eq:5}
\begin{array}{c}
(\sum^K_k{W_{kit}}\times u_{kit}+ \ {WO}_i\times {UO}_{it}) /\ (\sum^K_k{W_{kit}}+{WO}_i)        \\+\left(\sum^M_m{\sum^L_l{WC_{ijmlt}\times \ x_{ijml}}}\right)  \times \left(p_{ijt}\right) \le 1\\ \forall \ t, (i,j)\in{N}\ 
\end{array}
\end{equation}

The objective function is made up of 3 terms, all summed together. The construction cost of the charging pads (first term). Fixed and variable costs of wireless charging (second and third terms). The first constraint defined the average charging flows for each route between two intersections (see Eq. (\ref{eq:2})). The second constraint is to make sure the average charging flow for all tours is more than the minimum acceptable charging flow (see Eq. (\ref{eq:3})).Eqs. (\ref{eq:4}-\ref{eq:4.5}) are technical constraints to make sure only one setting is registered for the optimal location profile. The last constraint (see Eq. (\ref{eq:5})) is to make sure there is no additional charge from the charging pads when the average charging flow is 100\% (fully charged).
All the mentioned constraints in the deterministic model are summarized in Table \ref{tab:constraints}. 

\begin{longtable}{p{0.4in}p{0.45in}p{1.5in}} 
\caption{Constraints descriptions in the proposed model}
  \label{tab:constraints}\\ 
 \hline\hline 
 \endfirsthead
  \hline
 \multicolumn{3}{c}{Continuation of Table \ref{tab:constraints}}\\
 \hline
 \endhead
 \hline
  \multicolumn{3}{c}{Continued on next page}\\
  \hline
 \endfoot
  \hline
 \multicolumn{3}{c}{End of Table}\\
 \hline\hline
 \endlastfoot

\hline
\textbf{Title}                                                           &           \multicolumn{1}{c}{\textbf{Equation}}             & \multicolumn{1}{c}{\textbf{Remarks}} 

                                                                \\ \hline
\multicolumn{1}{l|}{\multirow{1}{*}{Average charging flow}}                        & \multicolumn{1}{c|}{(\ref{eq:2})}                    & The average charge of vehicles in a route is calculated by a weighted sum of the average charge of input routes, plus wireless charging, minus electricity consumption for that route.                              \\
                     
\multicolumn{1}{l|}{\multirow{1}{*}{Minimum acceptable charging level}}                  & \multicolumn{1}{c|}{(\ref{eq:3})}                    & Minimum charging levels should be larger than a determined value.\\
\multicolumn{1}{l|}{\multirow{1}{*}{Unique setting for wireless charging pads}}               & \multicolumn{1}{c|}{(\ref{eq:4}) and (\ref{eq:4.5})}                   & Between two nodes, we can’t have multiple settings for the wireless charging pads.                                \\
\multicolumn{1}{l|}{\multirow{1}{*}{State of charging pads}}               & \multicolumn{1}{c|}{(\ref{eq:5})}                   & If the average charging level is equal to 100\% then the charging pads should be off.                                \\

\end{longtable}

\subsection{Model Transformation}
The current model is not linear, and in a large-sized or a more detailed version of this problem, to analyze or find near-optimal decisions, we need to have a linear model.

The objective function in Eq. (\ref{eq:1}) and the constraint in Eq. (\ref{eq:5}) we face non-linearity.
For the linearization of the product of binary variables we refer interested readers to see \cite{adams2005simple,Glover1975IMPROVEDLI,or1}. Here we use the extensions of those concepts, and for example, to linearize the constraint in Eq. (\ref{eq:5}), we define a new supplementary variable as $z_{ijt}$ shown in Eq. (\ref{eq:7}) and transform the non-linear constraint to linear constraints using Eqs. (\ref{eq:8}-\ref{eq:11}).  

\begin{equation} \label{eq:7}
\begin{array}{c}
\left(\sum^M_m{\sum^L_l{WC_{ijmlt}\times \ x_{ijml}}}\right)\times \left(p_{ijt}\right)=\ z_{ijt}\ \ \ \ \ \forall \ t, (i,j)\in{N}\
\end{array}
\end{equation}

\begin{equation} \label{eq:8}
\begin{array}{c}
z_{ijt}\le \left(\sum^M_m{\sum^L_l{WC_{ijmlt}\times \ x_{ijml}}}\right)\ \forall \ t, (i,j)\in{N}\
\end{array}
\end{equation}

\begin{equation} \label{eq:9}
\begin{array}{c}
z_{ijt}\le M\times p_{ijt}\ \forall \ t, (i,j)\in{N}\
\end{array}
\end{equation}

\begin{equation} \label{eq:10}
\begin{array}{c}
z_{ijt}\ge \left(\sum^M_m{\sum^L_l{WC_{ijmlt}\times \ x_{ijml}}}\right)-\left(1-p_{ijt}\right)\times M\ \ \ \ \ \ \ \ \forall \ t, (i,j)\in{N}\
\end{array}
\end{equation}

\begin{equation} \label{eq:11}
\begin{array}{c}
z_{ijt}\ge 0\ \forall \ i,j,t\
\end{array}
\end{equation}

The final model is mixed-integer, multi-period, and linear. These features make the model more flexible and provide better tools to analyze different aspects of DWPT systems. 
\subsection{ Renewable Energy Integration}
One of the primary goals of developing EVs is to decrease the dependency on traditional fuels such as gasoline. As a result, renewable energy integration is one of the fundamental topics in DWPT transfer systems. The detailed modeling of renewable energy and smart grid integration is out of the scope of this research and is one of our focuses for future research. In an integrated planning decision model, power generated from the grid and solar energy will be added as variables (see Eq. (\ref{eq:12})) used in wireless charging infrastructures.
This research provides a less complex case analysis to show how solar energy can affect the system's performance. We assume the power as a  parameter. 

\begin{equation} \label{eq:12}
\begin{array}{c}
w_{ijt}\le g_{ijt}+s_{ijt}\ \ \ \ \ \ \ \forall \ t, (i,j)\in{N}\
\end{array}
\end{equation}

In the next section, a numerical example is illustrated to describe the model's performance and examine the effect of renewable energy on the model's reliability. 
\subsection{Numerical Example}
\begin{figure}
  \centering {
  
  \centering
\includegraphics[width=4.5in, height=3in]{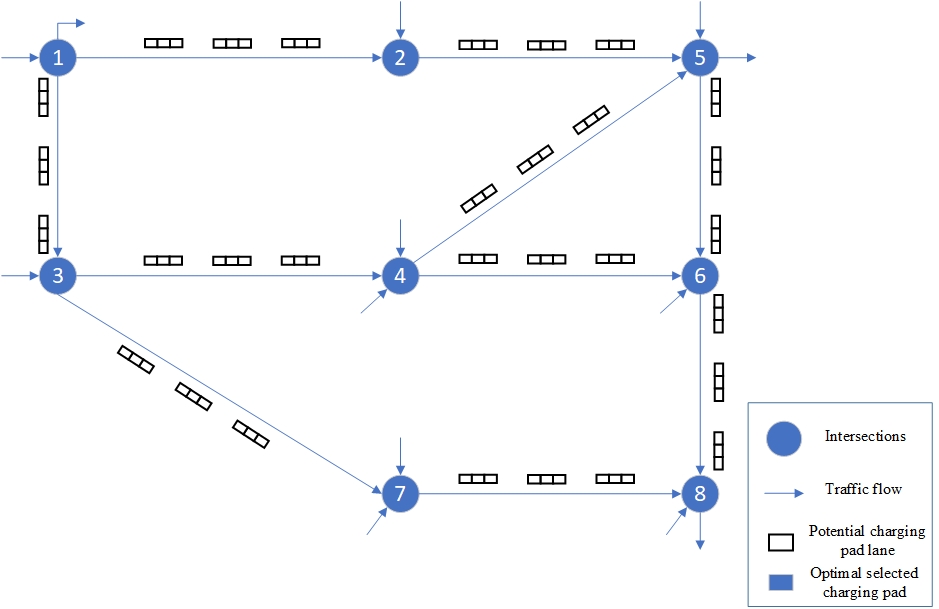}} 
  
  \caption{Network illustration of the connected roads}
  \label{fig1}
\end{figure}

\begin{figure}
  \centering {
  
  \centering
\includegraphics[width=4.5in, height=3in]{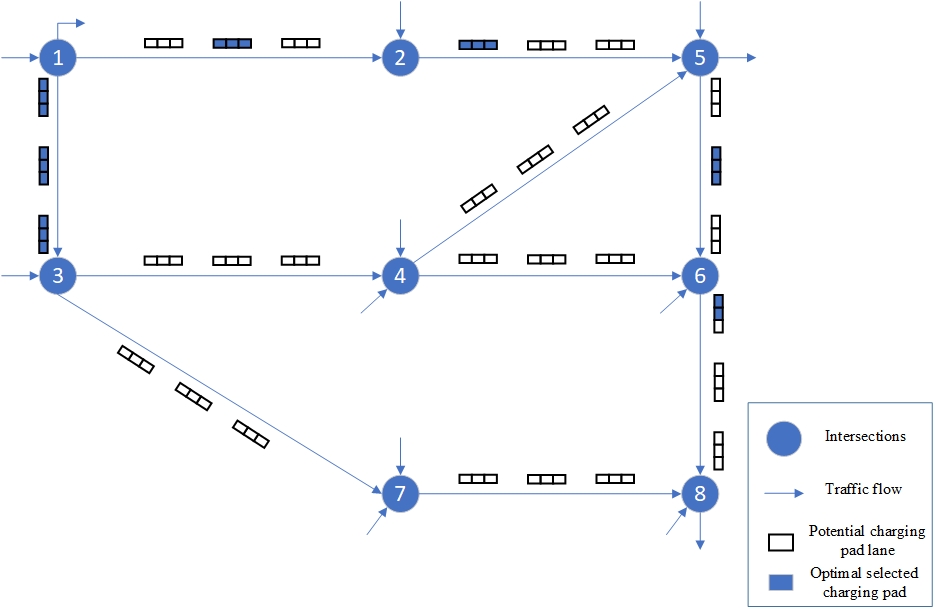}} 
  
  \caption{Network solution for the numerical example}
  \label{fig2}
\end{figure}

In this section, we introduce a numerical case study to analyze the decision generated by the proposed model. The data profile for the case study is presented in Table \ref{tab:example}. The network of the case study is illustrated in Fig. \ref{fig1}. The blue nodes in the network represent the intersections. The blue arrows between nodes are to show the route between intersections. The one-sided blue arrows enter the nodes are to determine the traffic flow. The big rectangular shapes (three small stacked rectangles) show the potential locations for wireless charging infrastructures. Each wireless charging location can have three different possible lengths. The larger the charging pads, the more charging capacity they can provide, and the more cost is needed to build them.
The model optimal decisions for wireless charging pads settings and locations are shown in Fig \ref{fig2}.The blue-filled rectangles determines the optimal selected locations and settings for the wireless charging infrastructures with a minimal total system cost. 
\begin{center}
\begin{longtable}[c]{p{0.9in}|p{3.5in}} 
\caption{Numerical example data profile}
\label{tab:example}\\ 
 \hline\hline 
 \endfirsthead
  \hline
  \multicolumn{2}{c}{Continuation of Table \ref{tab:example}}\\
 \endhead
  \multicolumn{2}{c}{Continued on next page}\\
  \hline
 \endfoot
  \hline
 \multicolumn{2}{c}{End of Table}\\
 \hline\hline
 \endlastfoot

\textbf{Sets} & \textbf{Description} \\ \hline 
\textit{I} & Set of nodes/intersections that \textit{i}= $\mathrm{\{}$1, 2,{\dots}, \textit{8}$\mathrm{\}}$ \\ \hline 
\textit{J} & Set of nodes/intersections that \textit{j}= $\mathrm{\{}$1, 2,{\dots}, \textit{8}$\mathrm{\}}$ \\ \hline 
\textit{T} & Hourly planning horizon that \textit{t}= $\mathrm{\{}$1, 2,{\dots}, \textit{24*30}$\mathrm{\}}$ \\ \hline 
\textit{M} & Set of potential locations for charging pads \textit{m}= $\mathrm{\{}$1, 2, \textit{3}$\mathrm{\}}$ \\ \hline 
\textit{L} & Set of potential lengths for charging pads \textit{l}= $\mathrm{\{}$1, 2, 3$\mathrm{\}}$ \\ \hline 
\textbf{Parameters} &  \\ \hline 
${CC}_{ijt}$ & Uniform (0.06,0.12) \\ \hline 
${CCV}_{ijmt}$ & Uniform (10,14)  \\ \hline 
${CCF}_{ijmt}$ & Uniform (100,110) \\ \hline 
${WE}_{ijt}$ & 1 \\ \hline 
${CCC}_{ijml}$ & 10000 \\ \hline 
$W_{ijt}$ & W('1','2')=3;\newline W('2','5')=5;\newline W('1','3')=7;\newline W('3','7')=5;\newline W('3','4')=3;\newline W('4','5')=8;\newline W('4','6')=4;\newline W('5','6')=4;\newline W('6','8')=7;\newline W('7','8')=8; \\ \hline 
$WC_{ijmlt}$ & Small = 0.04;\newline Medium =0.08;\newline Large =0.12; \\ \hline 
${UO}_{jt}$ & 0.5 \\ \hline 
${WO}_i$ & 2 \\ \hline 
$MA$ & 0.4 \\ \hline 
\textit{M} & A sufficiently large number \\ \hline 
\textbf{Variables} &  \\ \hline 
$x_{ijml}$ & Binary variable to indicate whether the potential location \textit{m} for route \textit{i to j} is selected for adding charging pads with the length of \textit{l}. \\ \hline 
$u_{ijt}$\textbf{} & Average charging level at route \textit{i} to \textit{j }at time \textit{t} \\ \hline 
$p_{ijt}$ & Binary variable to show the on and off state of the charging pads \\ \hline 

\end{longtable}
\end{center}
We also performed a sensitivity analysis to see the effect of minimum acceptable average charging flow on the decision model (see Table. \ref{tab:costcomparison}).

\begin{table}[]
\caption{Total system cost comparisons for different minimum acceptable charging flow}
\label{tab:costcomparison} 
\begin{adjustbox}{width=\textwidth}
\begin{tabular}{lcccccccccc}
\label{tab:costcomparison} 

\multirow{2}{*}{$ MA = \alpha$} & \multicolumn{9}{c}{Minimum acceptable average charging flow for EVs}                                                           \\ \cline{2-10} 
                                    & $\alpha = 0.2$      & $\alpha = 0.27$   & $\alpha = 0.3$      & $\alpha = 0.4$   & $\alpha = 0.5$      & $\alpha = 0.6$   & $\alpha = 0.7$      & $\alpha = 0.74$   & $\alpha = 0.8$         \\ \hline
Total Cost(\$)                & 85819 & 233003
    & 259027 & 606042
 & 865592 & 1212526
 & 1.56E+06
 & 1.56E+06
 & Infeasible
  \\
\hline
\end{tabular}
\end{adjustbox}
\end{table}

When the minimum average flow requirement \textit{MA} = 0.2, the decision model suggests building no wireless charging infrastructure; at \textit{MA} = 0.27, the first location for charging infrastructure is selected. The total system cost increases with the increase of our expectations for the \textit{MA}. At \textit{MA} = 0.74, the decision model proposes sixteen different sites with maximum lengths in the network for wireless charging infrastructure locations. If we have \textit{MA} larger than 0.74, even if the model increases the number of wireless charging locations due to the high expectations for minimum average charging flow, the model won't find any feasible solutions. The results also further aid in verifying the model's performance, as we expected to see an increase in the total system costs and an increase in the total selected locations to satisfy the minimum average charging flow requirements.  

\begin{figure}
  \centering {
  \centering
\includegraphics[width=4.55in, height=2.1in]{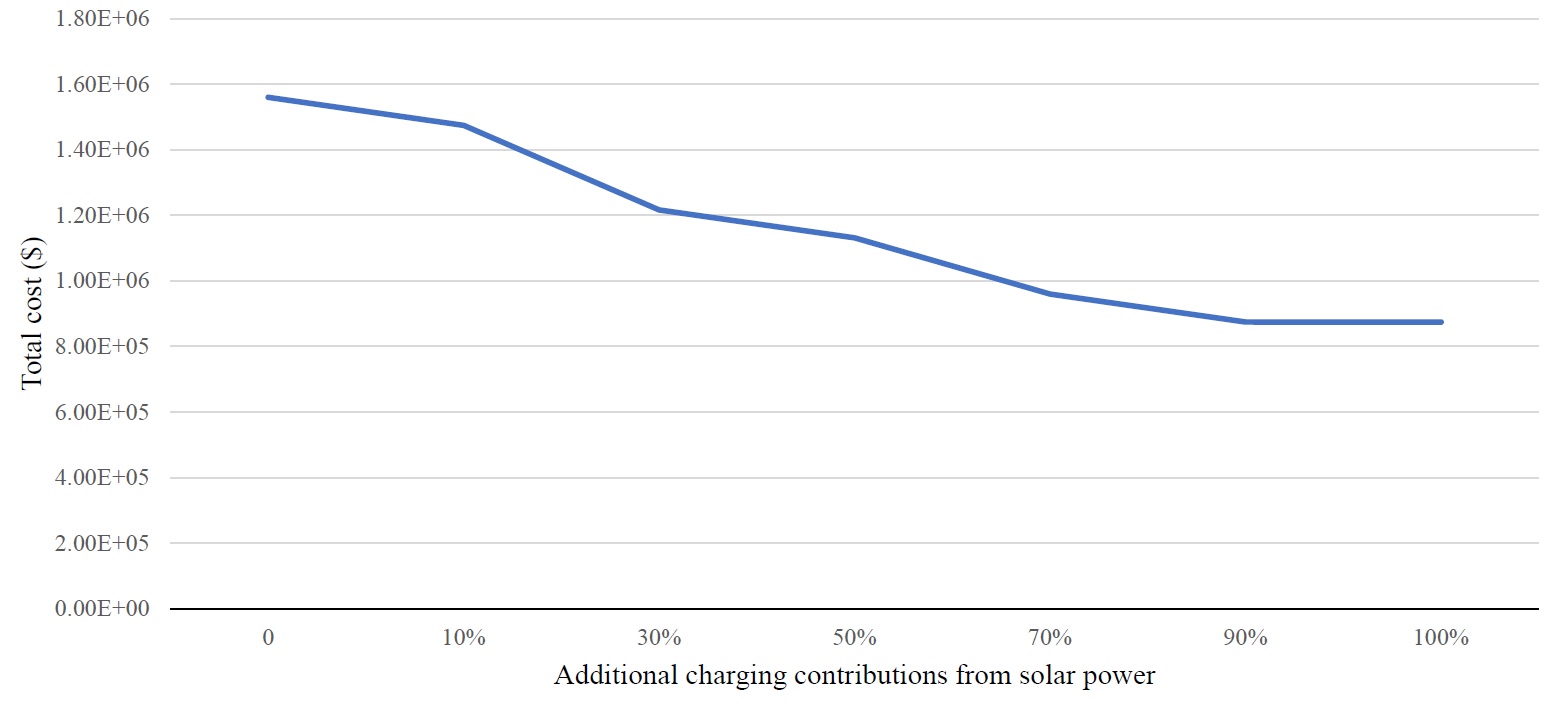}} 
  \caption{Total system cost and the solar power contribution on wireless charging }
  \label{ex3rutgers}
\end{figure}

To analyze the effect of renewable energy on the system's performance, we assume that the charge provided by the wireless charging infrastructures is equal to the limited power generated by the power grid and additional solar power available to each charging location. We assume that we already considered the cost of the solar infrastructures (e.g., batteries) in the construction cost. In the more detailed version of this model, we need to include the uncertainty surrounding solar power and a proper function for the battery performance. 
The rest of data profile is the same as Table \ref{tab:example} with the change of \textit{MA} = 0.7. We performed a set of sensitivity analyses on the additional potential contributions from solar power to wireless charging locations. The analysis results are presented in Table \ref{tab:ex3}, and Fig. \ref{ex3rutgers}.
\\

\begin{table}[]
\caption{Impact of the solar power integration on total system cost and optimal locations}
\label{tab:ex3} 
\begin{adjustbox}{width=\textwidth}
\begin{tabular}{@{}cccccccc@{}}
\toprule
Additional solar power contribution   & 0        & 10\%     & 30\%     & 50\%     & 70\%     & 90\%     & 100\%    \\ \midrule
Total system cost (\$)                & 1.56E+06 & 1.47E+06 & 1.22E+06 & 1.13E+06 & 9.59E+05 & 8.74E+05 & 8.74E+05 \\
Percentage decrease of the total cost & 0.00\%   & 5.48\%   & 22.04\%  & 27.47\%  & 38.49\%  & 43.94\%  & 43.97\%  \\
Selected locations with a small length  & 0        & 0        & 0        & 0        & 1        & 0        & 2        \\
Selected locations with a medium length & 0        & 1        & 0        & 1        & 1        & 2        & 0        \\
Selected locations with a large length  & 18       & 16       & 14       & 12       & 9        & 8        & 8        \\ \bottomrule
\end{tabular}
\end{adjustbox}
\end{table}

It is observed that the total system cost decreased as the potential contribution for solar power increased. At 0\% contributions from solar energy, a limited amount of power from the grid can contribute to the wireless power generation at each potential location. As a result, in the optimal decision, we see 18 different sites with the maximum length are selected to satisfy the minimum required average charging flow for the EVs. With the increase of the contribution from solar power, the total system cost decreases, and the number of locations needed for wireless charging pads decreases as well. At 50\% additional charging power contributions from solar energy, the number of sites selected in the optimal decision is reduced to 13 (12 with the maximum size and 1 with medium size). The total system cost also decreased by 27.47\%. That means the system needs fewer charging infrastructures to satisfy the EVs' energy demands. At 100\% additional contributions from solar energy, we have up to 43.97\% decrease in the total system cost. The number of optimal locations needed to satisfy EVs' demand also decreased from 18 to 10 (8 with the maximum size and 2 with the small size), a 44\% reduction compared to the initial planning at 0\% contribution from solar power. In the next section, we summarize and discuss the result of this research.

\section{Discussion}
Flourishing the electric vehicles market, the emergence of batteries with higher capacity with lower weights, and autonomous electric vehicles necessitate better DWPT systems and decision planning. To have a reliable decision model for DWPT systems, the interdisciplinary nature of the agents involved should be carefully analyzed. Otherwise, the generated decisions may even become infeasible in a real-world situation. 

In this study, after a brief review of the DWPT systems, we proposed a novel mathematical model to find the optimal locations and settings for wireless charging infrastructure. The advantage of this data-driven model is that it can work with available data such as "Average Annual Daily Traffics". The proposed model also has been developed to be connected to a systems-level smart grid and renewable energy decision model. The results indicate the high sensitivity of optimal location decisions to other factors such as minimum expectations for average charging flow. 

We also investigated the impact of renewable energy integration (solar power) on wireless charging systems when the available power from power grids for wireless charging infrastructures is limited. The analyzes indicate that solar power integration can significantly reduce the total system cost and reduce the number of locations needed for wireless charging infrastructures. The results are beneficial for areas with a higher density of infrastructures, such as highly populated urban areas, when the available power for wireless charging from the power grid is limited.  

To our knowledge, this is one of the first studies at a systems level that aims to connect different sub-systems of DWPT. This deterministic model will be used as a foundation for a more advanced approach such as the Robust optimization approach introduced in\cite{doi:10.1061/(ASCE)WR.1943-5452.0000757,pej,ghassemi2019system}, and the Fuzzy logic approach proposed in \cite{10.1007/978-3-030-66501-2_56,10.1007/978-3-030-66501-2_9}. The power grid and renewable energy integration with a more detailed model are also our future focus for continuing this research.

\bibliographystyle{splncs03_unsrt}
\bibliography{author}

\begin{thebibliography}{10}
\providecommand{\url}[1]{\texttt{#1}}
\providecommand{\urlprefix}{URL }

\bibitem{EIA2021}
EIA: {Monthly energy review (May)}. Tech. rep. (2021)

\bibitem{article2}
Requia, W., Mohamed, M., Higgins, C., Arain, A., Ferguson, M.: How clean are
  electric vehicles? evidence-based review of the effects of electric mobility
  on air pollutants, greenhouse gas emissions and human health. Atmospheric
  Environment (1967)  185,  64--77 (05 2018)

\bibitem{8353730}
Jeong, S., Jang, Y.J., Kum, D., Lee, M.S.: Charging automation for electric
  vehicles: Is a smaller battery good for the wireless charging electric
  vehicles? IEEE Transactions on Automation Science and Engineering  16(1),
  486--497 (2019)

\bibitem{HE2019452}
He, Y., Kockelman, K.M., Perrine, K.A.: Optimal locations of u.s. fast charging
  stations for long-distance trip completion by battery electric vehicles.
  Journal of Cleaner Production  214,  452--461 (2019),
  \url{https://www.sciencedirect.com/science/article/pii/S0959652618339040}

\bibitem{bansal_2015}
Bansal, P.: Charging of electric vehicle: Technology and policy implications.
  Journal of Science Policy \& Governance  6(1) (Feb 2015)

\bibitem{wevj9010009}
Domingues-Olavarría, G., Márquez-Fernández, F.J., Fyhr, P., Reinap, A.,
  Alaküla, M.: Electric roads: Analyzing the societal cost of electrifying all
  danish road transport. World Electric Vehicle Journal  9(1) (2018),
  \url{https://www.mdpi.com/2032-6653/9/1/9}

\bibitem{Song16}
Song, K., Koh, K.E., Zhu, C., Jiang, J., Wang, C., Huang, X.: A review of
  dynamic wireless power transfer for in‐motion electric vehicles. In: Coca,
  E. (ed.) Wireless Power Transfer, chap.~6. IntechOpen, Rijeka (2016),
  \url{https://doi.org/10.5772/64331}

\bibitem{su13031270}
Kwag, S.I., Hur, U., Ko, Y.D.: Sustainable electric personal mobility: The
  design of a wireless charging infrastructure for urban tourism.
  Sustainability  13(3) (2021), \url{https://www.mdpi.com/2071-1050/13/3/1270}

\bibitem{7878666}
Mohamed, A.A.S., Lashway, C.R., Mohammed, O.: Modeling and feasibility analysis
  of quasi-dynamic wpt system for ev applications. IEEE Transactions on
  Transportation Electrification  3(2),  343--353 (2017)

\bibitem{7731966}
Cirimele, V., Freschi, F., Mitolo, M.: Inductive power transfer for automotive
  applications: State-of-the-art and future trends. In: 2016 IEEE Industry
  Applications Society Annual Meeting. pp. 1--8 (2016)

\bibitem{JANG2018844}
Jang, Y.J.: Survey of the operation and system study on wireless charging
  electric vehicle systems. Transportation Research Part C: Emerging
  Technologies  95,  844--866 (2018),
  \url{https://www.sciencedirect.com/science/article/pii/S0968090X18304649}

\bibitem{9681af48fb114fe3b70035c6940db613}
Yilmaz, M., Krein, P.: Review of battery charger topologies, charging power
  levels, and infrastructure for plug-in electric and hybrid vehicles. IEEE
  Transactions on Power Electronics  28(5),  2151--2169 (Jan 2013)

\bibitem{7110445}
Miller, J.M., Jones, P., Li, J.M., Onar, O.C.: Ornl experience and challenges
  facing dynamic wireless power charging of ev's. IEEE Circuits and Systems
  Magazine  15(2),  40--53 (2015)

\bibitem{5709980}
Imura, T., Hori, Y.: Maximizing air gap and efficiency of magnetic resonant
  coupling for wireless power transfer using equivalent circuit and neumann
  formula. IEEE Transactions on Industrial Electronics  58(10),  4746--4752
  (2011)

\bibitem{8063878}
Tavakoli, R., Pantic, Z.: Analysis, design, and demonstration of a 25-kw
  dynamic wireless charging system for roadway electric vehicles. IEEE Journal
  of Emerging and Selected Topics in Power Electronics  6(3),  1378--1393
  (2018)

\bibitem{mohamadi2020nash}
Mohamadi, M., Bahrini, A.: A nash--stackelberg equilibrium model for internet
  and network service providers in the demand market—a scenario-based
  approach. Wireless Networks  26(1),  449--461 (2020)

\bibitem{8930076}
Mohamed, A.A.S., Zhu, L., Meintz, A., Wood, E.: Planning optimization for
  inductively charged on-demand automated electric shuttles project at
  greenville, south carolina. IEEE Transactions on Industry Applications
  56(2),  1010--1020 (2020)

\bibitem{Brecher2014ReviewAE}
Brecher, A., Arthur, D.: Review and evaluation of wireless power transfer (wpt)
  for electric transit applications (2014)

\bibitem{osti_1399989}
Onar, O.C., Foote, A.P.: A review of high-power wireless power transfer  (8
  2017), \url{https://www.osti.gov/biblio/1399989}

\bibitem{Gil2014ALR}
Gil, A., Taiber, J.: A literature review in dynamic wireless power transfer for
  electric vehicles: Technology and infrastructure integration challenges
  (2014)

\bibitem{doi:10.1177/0020294019885158}
Nayagam, V.S., Premalatha, L.: Optimization of power losses in electric vehicle
  battery by wireless charging method with consideration of the laser optic
  effect. Measurement and Control  53(3-4),  441--453 (2020),
  \url{https://doi.org/10.1177/0020294019885158}

\bibitem{9177076}
Yao, Y., Gao, S., Wang, Y., Liu, X., Zhang, X., Xu, D.: Design and optimization
  of an electric vehicle wireless charging system using interleaved boost
  converter and flat solenoid coupler. IEEE Transactions on Power Electronics
  36(4),  3894--3908 (2021)

\bibitem{9219241}
Luo, Z., Wei, X., Pearce, M.G.S., Covic, G.A.: Multiobjective optimization of
  inductive power transfer double-d pads for electric vehicles. IEEE
  Transactions on Power Electronics  36(5),  5135--5146 (2021)

\bibitem{Ahmad2018ACR}
Ahmad, A., Alam, M.S., Chabaan, R.: A comprehensive review of wireless charging
  technologies for electric vehicles. IEEE Transactions on Transportation
  Electrification  4,  38--63 (2018)

\bibitem{Majhi2020ASR}
Majhi, R.C., Ranjitkar, P., Sheng, M., Covic, G.: A systematic review of
  dynamic wireless charging infrastructure location problem for electric
  vehicles (2020)

\bibitem{Machura2019ACR}
Machura, P., Li, Q.: A critical review on wireless charging for electric
  vehicles. Renewable \& Sustainable Energy Reviews  104,  209--234 (2019)

\bibitem{JANG2015222}
Jang, Y.J., Jeong, S., Ko, Y.D.: System optimization of the on-line electric
  vehicle operating in a closed environment. Computers \& Industrial
  Engineering  80,  222--235 (2015),
  \url{https://www.sciencedirect.com/science/article/pii/S036083521400429X}

\bibitem{KO201567}
Ko, Y.D., Jang, Y.J., Lee, M.S.: The optimal economic design of the wireless
  powered intelligent transportation system using genetic algorithm considering
  nonlinear cost function. Computers \& Industrial Engineering  89,  67--79
  (2015),
  \url{https://www.sciencedirect.com/science/article/pii/S0360835215001710},
  maritime logistics and transportation intelligence

\bibitem{SUN2020102331}
Sun, X., Chen, Z., Yin, Y.: Integrated planning of static and dynamic charging
  infrastructure for electric vehicles. Transportation Research Part D:
  Transport and Environment  83,  102331 (2020),
  \url{https://www.sciencedirect.com/science/article/pii/S1361920919304146}

\bibitem{CHEN2016344}
Chen, Z., He, F., Yin, Y.: Optimal deployment of charging lanes for electric
  vehicles in transportation networks. Transportation Research Part B:
  Methodological  91,  344--365 (2016),
  \url{https://www.sciencedirect.com/science/article/pii/S0191261516303319}

\bibitem{CHEN2017185}
Chen, Z., Liu, W., Yin, Y.: Deployment of stationary and dynamic charging
  infrastructure for electric vehicles along traffic corridors. Transportation
  Research Part C: Emerging Technologies  77,  185--206 (2017),
  \url{https://www.sciencedirect.com/science/article/pii/S0968090X17300347}

\bibitem{7837718}
Manshadi, S.D., Khodayar, M.E., Abdelghany, K., Üster, H.: Wireless charging
  of electric vehicles in electricity and transportation networks. IEEE
  Transactions on Smart Grid  9(5),  4503--4512 (2018)

\bibitem{NGO2020102385}
Ngo, H., Kumar, A., Mishra, S.: Optimal positioning of dynamic wireless
  charging infrastructure in a road network for battery electric vehicles.
  Transportation Research Part D: Transport and Environment  85,  102385
  (2020),
  \url{https://www.sciencedirect.com/science/article/pii/S1361920920305721}

\bibitem{FULLER2016343}
Fuller, M.: Wireless charging in california: Range, recharge, and vehicle
  electrification. Transportation Research Part C: Emerging Technologies  67,
  343--356 (2016),
  \url{https://www.sciencedirect.com/science/article/pii/S0968090X16000668}

\bibitem{article1}
He, F., Yin, Y.: Deploying public charging stations for electric vehicles on
  urban road networks. Transportation Research Part C: Emerging Technologies
  60,  227--240 (11 2015)

\bibitem{RePEc:eee:transb:v:103:y:2017:i:c:p:30-55}
Liu, H., Wang, D.Z.: {Locating multiple types of charging facilities for
  battery electric vehicles}. Transportation Research Part B: Methodological
  103(C),  30--55 (2017),
  \url{https://ideas.repec.org/a/eee/transb/v103y2017icp30-55.html}

\bibitem{su12218971}
Chen, G., Hu, D., Chien, S., Guo, L., Liu, M.: Optimizing wireless charging
  locations for battery electric bus transit with a genetic algorithm.
  Sustainability  12(21) (2020),
  \url{https://www.mdpi.com/2071-1050/12/21/8971}

\bibitem{article5}
Dong, J., Liu, C., Lin, Z.: Charging infrastructure planning for promoting
  battery electric vehicles: An activity-based approach using multiday travel
  data. Transportation Research Part C: Emerging Technologies  38,  44–55 (01
  2014)

\bibitem{article}
Zhang, A., Kang, J.E., Kwon, C.: Incorporating demand dynamics in multi-period
  capacitated fast-charging location planning for electric vehicles.
  Transportation Research Part B: Methodological  103 (05 2017)

\bibitem{Riemann2015OptimalLO}
Riemann, R., Wang, D.Z.W., Busch, F.: Optimal location of wireless charging
  facilities for electric vehicles: Flow-capturing location model with
  stochastic user equilibrium. Transportation Research Part C-emerging
  Technologies  58,  1--12 (2015)

\bibitem{SHI20201067}
Shi, R., Li, S., Zhang, P., Lee, K.Y.: Integration of renewable energy sources
  and electric vehicles in v2g network with adjustable robust optimization.
  Renewable Energy  153,  1067--1080 (2020),
  \url{https://www.sciencedirect.com/science/article/pii/S0960148120302135}

\bibitem{7815433}
Liu, S., Etemadi, A.H.: A dynamic stochastic optimization for recharging
  plug-in electric vehicles. IEEE Transactions on Smart Grid  9(5),  4154--4161
  (2018)

\bibitem{CAO2020105628}
Cao, Y., Huang, L., Li, Y., Jermsittiparsert, K., Ahmadi-Nezamabad, H.,
  Nojavan, S.: Optimal scheduling of electric vehicles aggregator under market
  price uncertainty using robust optimization technique. International Journal
  of Electrical Power \& Energy Systems  117,  105628 (2020),
  \url{https://www.sciencedirect.com/science/article/pii/S0142061518335518}

\bibitem{WANG2020119886}
Wang, Z., Jochem, P., Fichtner, W.: A scenario-based stochastic optimization
  model for charging scheduling of electric vehicles under uncertainties of
  vehicle availability and charging demand. Journal of Cleaner Production  254,
   119886 (2020),
  \url{https://www.sciencedirect.com/science/article/pii/S0959652619347560}

\bibitem{LIU201777}
Liu, Z., Song, Z.: Robust planning of dynamic wireless charging infrastructure
  for battery electric buses. Transportation Research Part C: Emerging
  Technologies  83,  77--103 (2017),
  \url{https://www.sciencedirect.com/science/article/pii/S0968090X17301985}

\bibitem{8402042}
Kosmanos, D., Maglaras, L.A., Mavrovouniotis, M., Moschoyiannis, S., Argyriou,
  A., Maglaras, A., Janicke, H.: Route optimization of electric vehicles based
  on dynamic wireless charging. IEEE Access  6,  42551--42565 (2018)

\bibitem{adams2005simple}
Adams, W.P., Forrester, R.J.: A simple recipe for concise mixed 0-1
  linearizations. Operations research letters  33(1),  55--61 (2005)

\bibitem{Glover1975IMPROVEDLI}
Glover, F.: Improved linear integer programming formulations of nonlinear
  integer problems. Management Science  22,  455--460 (1975)

\bibitem{or1}
Oral, M., Kettani, O.: A linearization procedure for quadratic and cubic
  mixed-integer problems. Operations Research  40(1-supplement-1),  S109--S116
  (1992), \url{https://doi.org/10.1287/opre.40.1.S109}

\bibitem{doi:10.1061/(ASCE)WR.1943-5452.0000757}
Ghassemi, A., Hu, M., Zhou, Z.: Robust planning decision model for an
  integrated water system. Journal of Water Resources Planning and Management
  143(5),  05017002 (2017)

\bibitem{pej}
Peykani, P., Mohammadi, E., Saen, R.F., Sadjadi, S.J., Rostamy-Malkhalifeh, M.:
  Data envelopment analysis and robust optimization: A review. Expert Systems
  37(4),  e12534 (2020),
  \url{https://onlinelibrary.wiley.com/doi/abs/10.1111/exsy.12534}

\bibitem{ghassemi2019system}
Ghassemi, A.: System of Systems Approach to Develop an Energy-Water Nexus Model
  Under Uncertainty. Ph.D. thesis (2019)

\bibitem{10.1007/978-3-030-66501-2_56}
Ghassemi, A., Scott, M.J.: Investigating the role of renewable energies in
  integrated energy-water nexus planning under uncertainty using fuzzy logic.
  In: Allahviranloo, T., Salahshour, S., Arica, N. (eds.) Progress in
  Intelligent Decision Science. pp. 696--702. Springer International
  Publishing, Cham (2021)

\bibitem{10.1007/978-3-030-66501-2_9}
Ghassemi, A., Scott, M.J.: A mathematical approach to improve energy-water
  nexus reliability using a novel multi-stage adjustable fuzzy robust approach.
  In: Allahviranloo, T., Salahshour, S., Arica, N. (eds.) Progress in
  Intelligent Decision Science. pp. 115--123. Springer International
  Publishing, Cham (2021)

\end{thebibliography}
\end{document}